\documentclass[twocolumn,showpacs,preprintnumbers,amsmath,amssymb]{revtex4}
\usepackage{graphicx}% Include figure files
\usepackage{dcolumn}% Align table columns on decimal point
\usepackage{bm}% bold math

\begin{document}

%\preprint{APS/123-QED}

\title{Removing the trend of drift induced from acceleration noise for LISA}

\author{Alf Tang}
\affiliation{Blackett Laboratory, Imperial College London, Prince Consort Road, London, SW7 2AZ}
\email{alf.tang@nspo.narl.org.tw}

\author{Timothy J. Sumner}
\affiliation{Blackett Laboratory, Imperial College London, Prince Consort Road, London, SW7 2AZ}
\email{t.sumner@imperial.ac.uk}

\date{January 12, 2012}

\begin{abstract}

In this paper we demonstrate a methodology to remove the power of the drift induced from random acceleration on LISA proof mass in the frequency domain.  The drift must be cleaned from LISA time series data in advance of any further analysis.  The cleaning is usually performed in the time domain by using a quadratic function to fit the time series data, and then removing the fitted part from the data.  Having Fourier transformed the residuals, and then convolved with LISA transfer function, LISA sensitivity curve can be obtained.  However, cosmic gravitational-wave background cannot be retrieved with this approach due to its random nature.  Here we provide a new representation of power spectrum given by discrete Fourier transform, which is applied to find the function of the drift power for the cleaning in the frequency domain.  We also give the probability distribution used to analyze the data in the frequency domain.  We combine several techniques, including Markov Chain Monte Carlo method, simulated annealing, and Gelman \& Rubin's method, with Baye's theorem to build the algorithm.  The algorithm is utilized to analyze 24 simulations of LISA instrumental noise.  We prove that the LISA sensitivity can be recovered through this approach.  It can help us to build algorithms for some tasks which are must accomplished in the frequency domain for LISA data analysis.  This method can be applied to other space-borne interferometers if charges on their proof masses cannot be perfectly cancelled.

\end{abstract}

\maketitle

\section{Introduction}

In the LISA data stream, intrinsic instrumental noise falls into two categories: shot noise and 
acceleration noise \cite{prephaseA}.  
Acceleration noise, caused by the residual Coulomb force induced from 
the imperfect cancellation of charges on proof-masses, is dominant in the low-frequency range, resulting the sensitivity proportional to $f^{-2}$ roughly below 2 mHz.  
Optical-path noise, including mainly shot noise and beam-pointing error, is dominant in the high frequency range, leading the sensitivity declining proportional to the frequency above 10 mHz due to the falloff of the antenna transfer function.  

The theoretical LISA sensitivity can be obtained from the various types of noise spectral densities directly \cite{prephaseA}, whereas when we deal with LISA time series raw data, the trend of the drift of proof mass in the time series due to the random acceleration shall be removed at first.  
In general, the removal is performed in the time domain.  
By Fourier transforming the residuals, the LISA sensitivity is recovered.  
However, if stochastic gravitational-wave background exists in the data, bias might be induced if the trend is firstly removed in the time domain and then the background is extracted in the frequency domain.  
The analysis given by such sequential subtraction might be inaccurate, 
particularly if signals have overlaps.  
For instance, if a data set containing two overlapped signals is fitted by a linear filter where the signal is parametrized by a  rectangular function of amplitude and location, the filter may extract a stronger output around the overlapped region instead of one of the exact signals.  
This is because what fitting does is to minimize the difference between data and model.  
For that reason, the parameter estimation and the removal of the trend in the LISA data analysis should all be performed either in the time domain or in the frequency domain. 

The cosmological sources in the very early universe are randomly distributed across the sky, 
emitting gravitational waves with various amplitude and frequency.  
If they are not strong enough to be located by LISA, their incoherent signals will form a 
continuum and be entangled with instrumental noise.  
In order to gain cosmological information, the functional form of the trend and 
cosmic gravitational-wave background (CGB) are both needed to be understood.

The waveform of CGB in the time domain cannot be obtained due to its stochastic nature, 
making the separation of CGB from the time series data very difficult.  
It is natural to extract the CGBs in the frequency domain since the function of the power spectra can be written analytically.  Nevertheless, the method to remove the drift in the frequency domain is unknown.  
In this paper, the function of the power spectrum of the drift will be derived, 
and a method to remove the trend in the frequency domain for LISA data analysis will be developed.

In section \ref{sec:time-domain}, we will review the time series of LISA instrumental 
noise, and demonstrate the approach conducted in the time domain to obtain the LISA sensitivity.  
In section \ref{sec:ins-noise-power}, we will find a new representation of Fourier power spectrum, and use the representation to derive the power spectrum of the drift trend.  
In section \ref{sec:noise-power-prob}, the derivation of probability distribution of noise power will be provided.  
In section \ref{sec:analysis}, the algorithm of parameter estimation will be introduced, followed by the result of data analysis.  
Finally, in section \ref{sec:conclusion}, conclusion will be given.

\section{Remove Displacement Caused by LISA Random Acceleration in the Time Domain}
\label{sec:time-domain}

\begin{figure}
\includegraphics[scale=0.31]{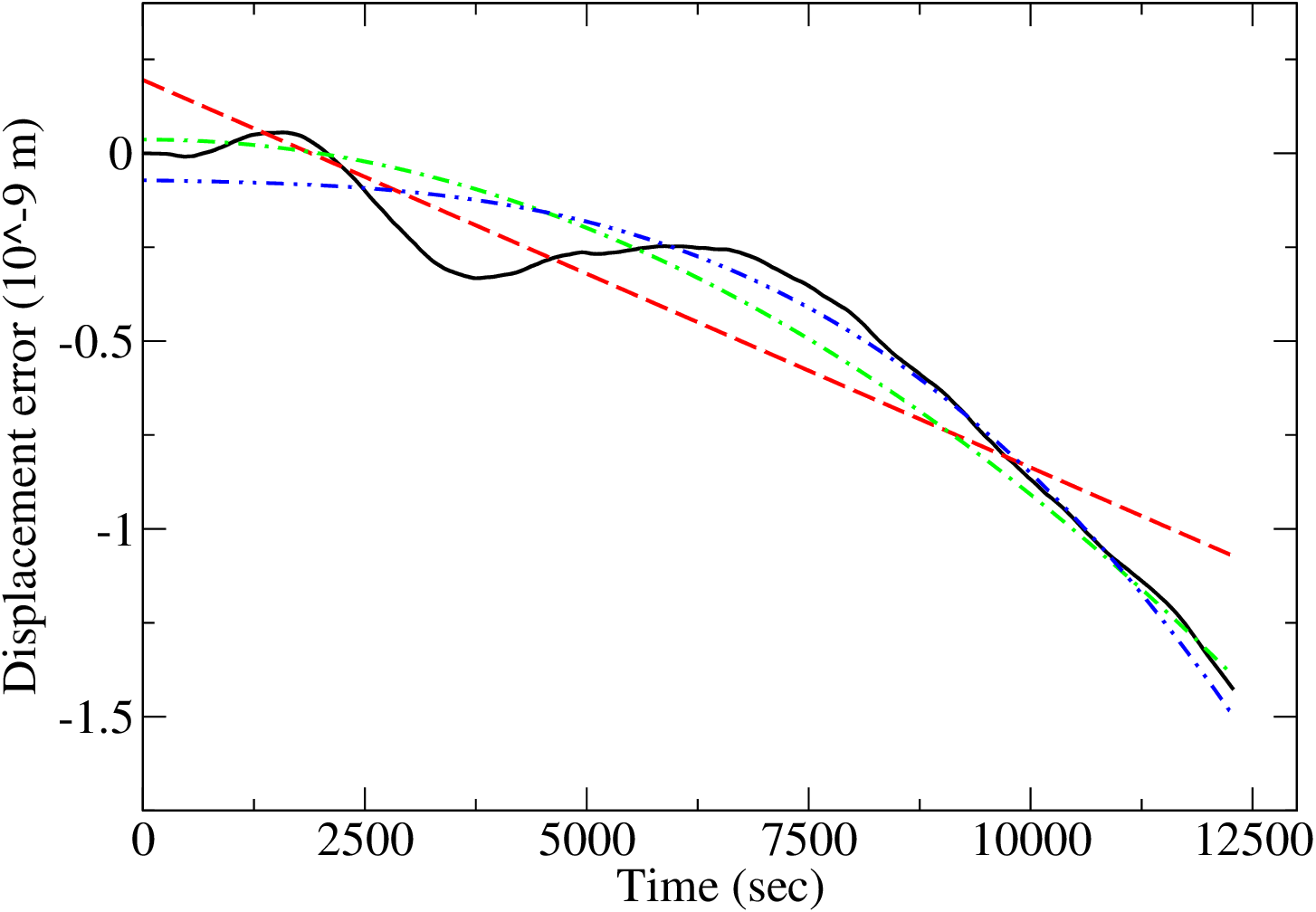}
\caption{\label{fig:time-fit} The simulation of displacement shift of proof mass due to random acceleration.  The black curve is simulation shifts.  The red line (dash) is the best fit given by a linear function.  The green curve (dash-dot) is the best fit given by a quadratic function.  The blue curve (dash-double-dot) is the best fit given by a third-power function.}
\end{figure}

Energetic particles keep hitting the proof mass, producing random accelerations on it continuously.  
The perturbations from random accelerations accumulate, and gradually depart the proof mass from its free-falling trajectory \cite{LISASIM}.  The LISA sensitivity curve cannot be obtained just by directly Fourier transforming the drift induced from such accelerations. 
The drift must be fitted by a quadratic function and the best fit must be removed.  
Then the Fourier transformed residuals can represent the LISA noise level.
In this section we will demonstrate how LISA sensitivity curve is obtained from fitting in the time domain.  Firstly we will simulate the drift induced from the acceleration, and then fit the simulated data.
We utilized Gaussian distribution to simulate the shifts \cite{LISASIM}.  
The acceleration noise power spectral density is suggested as $9.0\times 10^{-30}\ m^{2}/s^{4}Hz$ \cite{prephaseA}.  
Dividing it by sampling rate $dt$ and then taking square root, we will obtain the average of acceleration.  
Using this average as the standard deviations, we can draw a time series of random accelerations.  
Having double integrated the random accelerations, we will have the time series drift.  
Here the sampling rate $dt$ is set as 1.5 second, and 8192 data are simulated.  
The simulation of drift is shown by the black curve in 
Fig. \ref{fig:time-fit}.  

\begin{figure}
\includegraphics[scale=0.31]{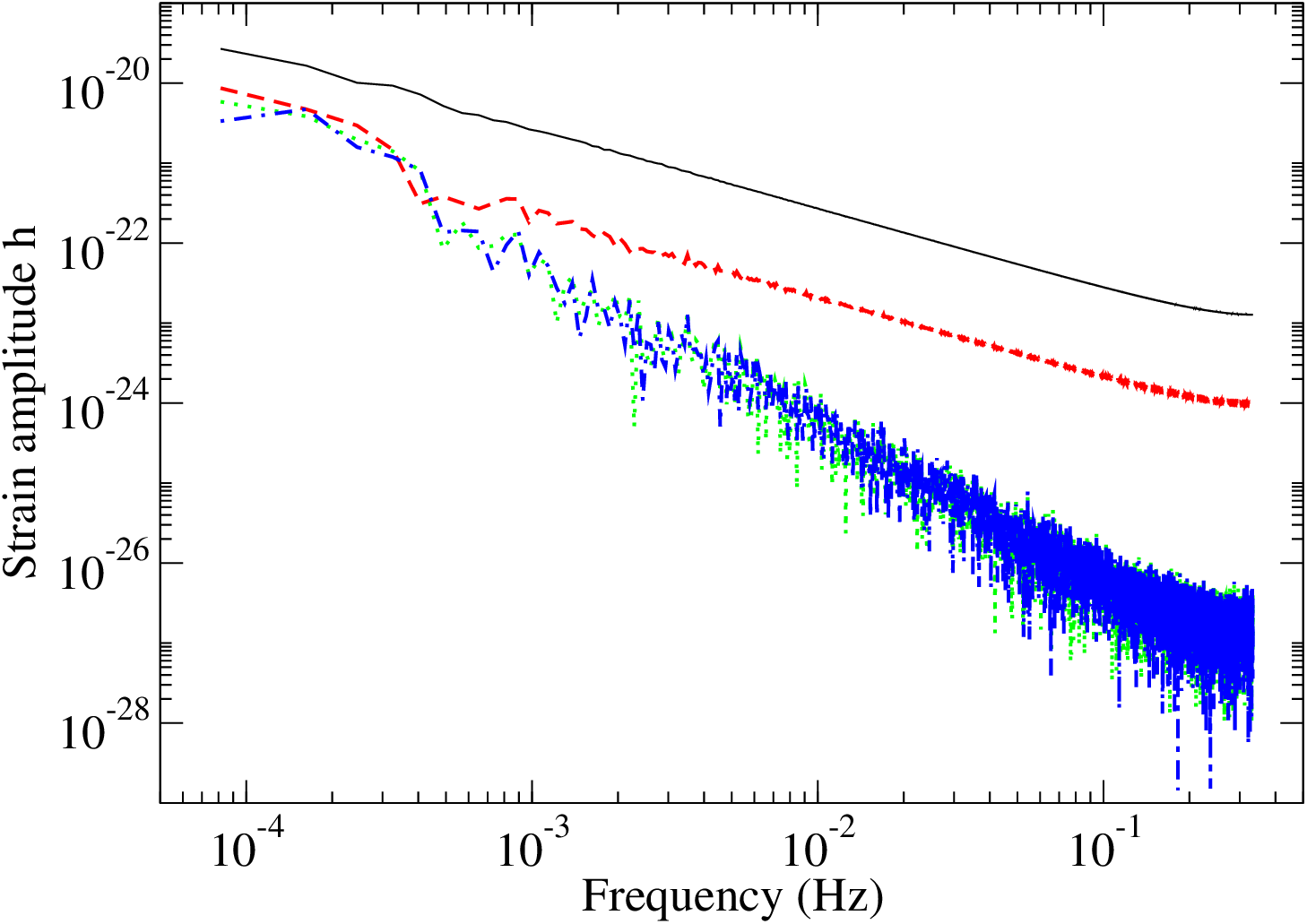}
\caption{\label{fig:residual-FTed} The black line is the displacement error introduced 
by the acceleration noise in the frequency domain.  The red dash line is the residual of 
linear fitting in the frequency domain.  The green line is the residual of quadratic fitting 
in the frequency domain.  The blue line is the residual of cubic fitting in the frequency domain.}
\end{figure}

Dividing the drift by arm-length %$L=5\times10^9$ m 
the dimensionless drift can be obtained.  
Having Fourier transformed the dimensionless drift, we will get the strain amplitude, 
as shown by the solid line in Fig. \ref{fig:residual-FTed}.  
As shown in the Fig.  \ref{fig:residual-FTed}, the frequency dependence of the strain is approximately $1/f$ rather than $1/f^{2}$ as indicated by the sensitivity curve in the LISA Prephase A study.  
In order to recover the LISA sensitivity the trend of drift must be removed.  

Since the drift is induced from random acceleration, it is natural to model its trend by a quadratic function $at^{2}+bt+c$ where a, b, c are unknown parameters.  
Using the function to fit the drift, and then remove the trend, which is identified as green dash-dot curve in the Fig. \ref{fig:time-fit}, from the drift, the residual can be obtained.  
By Fourier transforming the residual, the true noise level can be recovered.  
The strain amplitude of the residual is shown by the green dot curve 
in Fig. \ref{fig:residual-FTed}.  
Its frequency dependance is proportional to $f^{-2}$ 
as shown in the LISA Prephase A study \cite{prephaseA}.  

In addition to the quadratic equation, the linear equation $at+b$ and the cubic equation 
$at^{3}+bt^{2}+ct+d$ are tested to remove the trend as well.  
The best fits given by the linear and cubic equation are indicated by red and blue line, respectively.  
From Fig. \ref{fig:residual-FTed} it is noticed that the amplitude of the residual given 
by the fitting with linear function is lower than the uncleaned displacement by one order of magnitude, but it is inversely proportional to the frequency $f$, not to $f^{2}$.  
The amplitude of the residual given by the fitting with cubic function is still proportional to $f^{-2}$, and is as the same level of the amplitude corresponding to quadratic fitting.  
One more parameter used in the cubic fitting does not provide extra benefit. 

Here we demonstrated that the LISA sensitivity curve is obtained by removing 
the trend of the drift.  There is no problem if only point sources involve in data since point sources can be analyzed simultaneously with removing the trend in the time domain.  
However, if data contain the waves produced by stochastic sources, such as astrophysical foreground or cosmological background, which can only be analyzed in the frequency domain, 
removing the trend beforehand in the time domain would induce a bias in the analysis of the sources.  
A technique to deal with the trend in the frequency domain is necessary.

\section{Expected Power of Instrumental Noise}
\label{sec:ins-noise-power}

In last section the way to remove the trend of the drift in the time domain was described.  
In this section, a method to remove the trend directly in the frequency domain will be shown.

\begin{figure}
\includegraphics[scale=0.31]{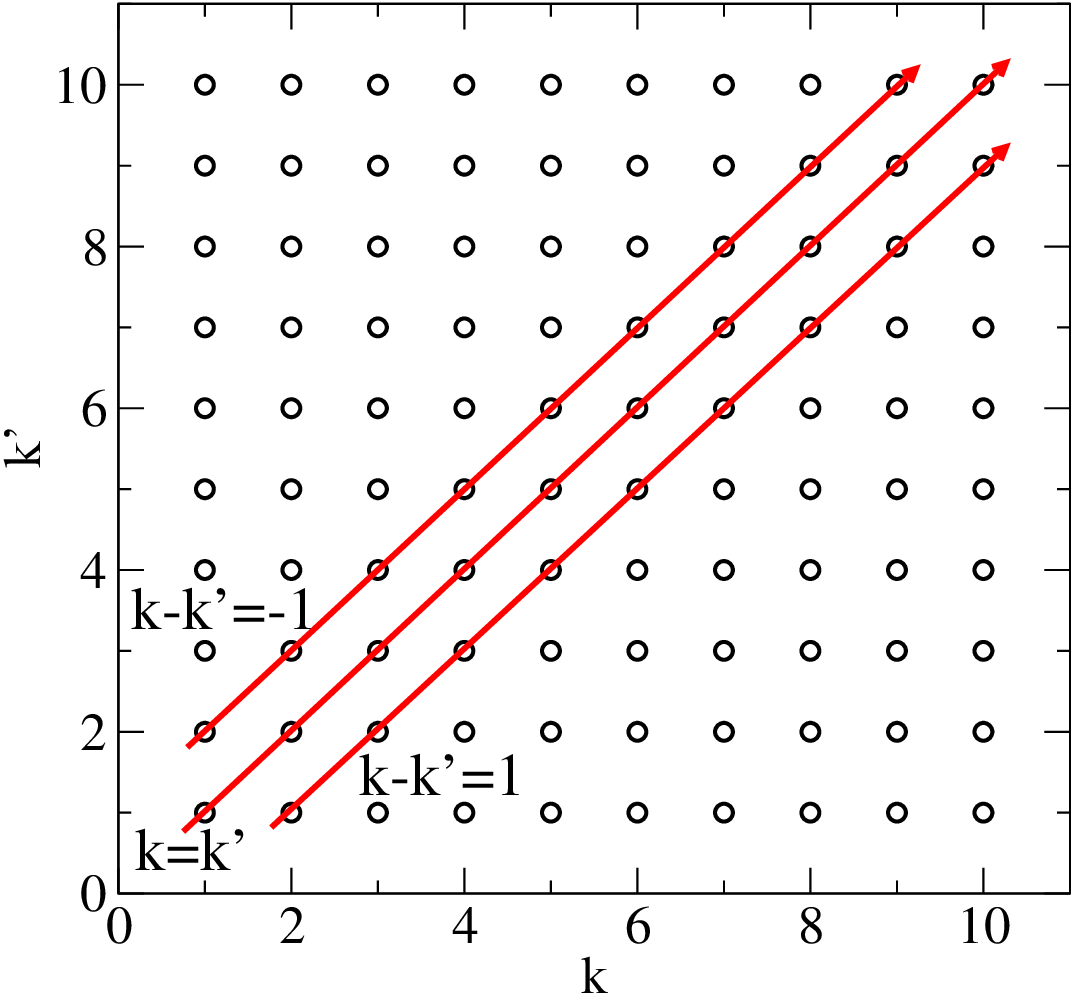}
\caption{\label{fig:grid} Illustration of the summation of Fourier power spectrum.}
\end{figure}

Suppose $h_{k}, k: 0 ~ N-1$ is the time series of data, and $H_{n}$ is the Fourier transform 
of the data, the power $| H_{n}|^{2}$ is given by
\begin{eqnarray}
| H_{n}|^{2} &=& H_{n} \times H_{n}^{*} \\
 &=& \sum_{k=0}^{N-1}\sum_{k'=0}^{N-1}h_{k}h_{k'}\exp\Big\{
 \frac{2\pi i(k-k')n}{N}\Big\},    
\label{eq:power}   
\end{eqnarray}
where $k$ is index, and $N$ is total number of data.
The summation is usually calculated firstly over one index and then the other, or vice versa.  
This is implied by the functionality of summation.  
However, what does matter is summing the term over all k and k' on the $N\times N$ grid.  
The implied procedure is not the only approach to carry out the calculation.  
We calculate the summation along the diagonal arrays as shown in the 
Figure \ref{fig:grid} rather than the regular procedure.  
The terms along red lines have the property that the difference of $k$ and $k'$ is fixed.  
Thus we introduce an index $t$ to indicate their difference $k-k'$.  
$k$ equals $k'$ on the diagonal line, so their phase is cancelled.  The term 
$h_{k}h_{k'}\exp\{2\pi in(k-k')/N\}$ turns to be $h^{2}_{k}$.  Considering the arrays 
corresponding to $t=1$ and $t=-1$, the terms along those diagonal arrays have 
symmetric mathematical expression.  
Their phases, $2\pi n/N$ and $-2\pi n/N$, have same magnitude but opposite sign.  
Because of that, the sum of those two is $2h_{k}h_{k+1}\cos(2\pi n/N)$ 
where $k:0\sim N-2$.
Similarly the other symmetric terms with $\pm t$ can be combined to 
$2h_{k}h_{k+t}\cos(2\pi nt/N)$ where $k:0\sim N-1-t$.  Therefore, we can rewrite 
Eq.\ \eqref{eq:power} as the following form
\begin{equation}
 | H_{n}|^{2} = \sum_{k=0}^{N-1} h_{k}^{2} + 2\sum_{t=1}^{N-1}\sum_{k=0}^{N-1-t}
        h_{k}h_{k+t}\cos \frac{2\pi nt}{N}. 
\label{eq:FFT}
\end{equation}

The benefit of new representation is that the time series data 
can be separated into autocorrelation terms $h^{2}_{k}$ and cross-correlation terms 
$h_{k}h_{k+t}$.  
When we deal with the Fourier component of random noise, 
the autocorrelation term will remain and cross-correlation term will vanish as their ensemble 
average is taken.   The ensemble average is associated with the standard deviation of 
the time series random noise.  As a conclusion, with this representation, the power spectrum of 
random noise can be expressed by some statistical properties of the time series data.

\subsection{Shot Noise}

Time series of shot noise can be characterised as independent Gaussian noise.  This implies 
that their ensemble average is zero, and they are not correlated, which can be expressed by 
$\langle h_{k}h_{k+t}\rangle=0$ for any $t\not= 0$.  The standard deviation of Gaussian 
distribution $\sigma^{2}$ should equal to the ensemble average of square of data 
$\langle h^{2}_{k}\rangle$ if data length is infinite long.  
Knowing this we can apply Eq. \eqref{eq:FFT} to find a formula for the Fourier 
component of shot noise.  Substituting time series shot noise $\{h_{k}\}$ into 
Eq. \eqref{eq:FFT} we will obtain
\begin{eqnarray}
\langle P_{n}\rangle &=& \langle |H_{n}|^{2}\rangle = 
\frac{1}{N}\sum_{k=0}^{N-1} \langle h_{k}^{2} \rangle \nonumber\\
&+& \frac{2}{N}\sum_{t=1}^{N-1}\sum_{k=0}^{N-1-t}
\langle h_{k}h_{k+t}\rangle\cos \frac{2\pi nt}{N}
\label{eq:ensemble}
\\
&=& \sigma^{2}.
\end{eqnarray}
It is not surprised that the formula for describing the power spectrum is a constant since shot noise is white noise.

\subsection{Acceleration Noise}

To find the expression for the power spectrum of acceleration noise, we review the 
characteristics of time series drift induced from random acceleration $\{a_{i}\}$ at first.  
The gross feature of $\{a_{i}\}$ can be recognised as independent Gaussian noise as well
\begin{equation}
P(a_{i}|I) = \frac{1}{\sqrt{2\pi}a}exp\left\{-\frac{a_{i}^{2}}{2a^{2}}\right\}.
\end{equation}  
First of all, the ensemble average $\langle a_{i}\rangle$ is zero.  
Secondly, the random acceleration noise $\{a_{i}\}$ is independent, so we have 
$\langle a_{i}a_{j}\rangle\ \forall i\not= j$.  Thirdly, the ensemble average of square 
of $\langle a^{2}_{i}\rangle$ equals to $a^{2}$ where $a$ is the standard deviation 
of the distribution.  

Now we construct the Fourier component of the displacement noise step by step.  We begin 
this work by establishing the Fourier component of velocity noise.
Suppose $v_{k}$ is the time series of velocity noise associated with random acceleration $a$ 
and its initial value $v_{0}$ is zero.  We assume that it obeys the equation  
\begin{equation}
v_{k}=v_{k-1}+a_{k}\Delta.
\end{equation}
With the initial condition we can derive
\begin{equation}
v_{k}=  (a_{1}+a_{2}+...+a_{k})\Delta,
\label{eq:hk}
\end{equation}
and
\begin{equation}
v_{k+t}=  (a_{1}+a_{2}+...+a_{k+t})\Delta.
\label{eq:hk+t}
\end{equation}

Then, we know
\begin{eqnarray}
v_{k}^{2} &=& \Delta^{2}\big(\sum_{i=1}^{k}a_{i}\big)
                         \big(\sum_{j=1}^{k}a_{j}\big)\\
                 &=& \Delta^{2}\big(\sum_{i=1}^{k}a_{i}^{2}
                        +2\sum_{i=1}^{k}\sum_{j=i+1}^{k}a_{i}a_{j}\big),     
\label{eq:hk2}                          
\end{eqnarray}
and
\begin{eqnarray}
v_{k}v_{k+t} &=& \Delta^{2}\big(\sum_{i=1}^{k}a_{i}\big)
                              \big(\sum_{j=1}^{k+t}a_{j}\big)\\
                      &=& \Delta^{2}\big(\sum_{i=1}^{k}a_{i}\big)
                              \big(\sum_{j=1}^{k}a_{j}+\sum_{j=k+1}^{k+t}a_{j}\big)\\
                      &=& \Delta^{2}\Big[\sum_{i=1}^{k}a_{i}^{2}
                      +2\sum_{i=1}^{k}\sum_{j=i+1}^{k}a_{i}a_{j}\nonumber\\
                      &+&\sum_{i=1}^{k}\sum_{j=k+1}^{k+t}a_{i}a_{j}\Big].      
\label{eq:hkhkt}        
\end{eqnarray}
Substituting Eq.\ \eqref{eq:hk2} and \eqref{eq:hkhkt} into Eq.\ \eqref{eq:FFT}, 
we can expand the power of velocity noise $|V_{n}|^{2}$ as
\begin{eqnarray}
|V_{n}|^{2} &=& \frac{\Delta^{2}}{N}\sum_{k=1}^{N-1}\sum_{i=1}^{k}a_{i}^{2}
                + 2\frac{\Delta^{2}}{N}\sum_{k=1}^{N-1}\sum_{i=1}^{k}\sum_{j=i+1}^{k}a_{i}a_{j}\nonumber\\
              &+& 2\frac{\Delta^{2}}{N}\sum_{t=1}^{N-2}\sum_{k=1}^{N-1-t}\cos \frac{2\pi nt}{N}
              \times \Big[\sum_{i=1}^{k}a_{i}^{2}\nonumber\\
              &+&2\sum_{i=1}^{k}\sum_{j=i+1}^{k}a_{i}a_{j}+\sum_{i=1}^{k}\sum_{j=k+1}^{k+t}a_{i}a_{j}\Big].
\label{eq:power1}
\end{eqnarray}
Then we can calculate the expected power of velocity noise by taking ensemble average of 
$|V_{n}|^{2}$
\begin{eqnarray}
\langle|V_{n}|^{2}\rangle &=& \frac{\Delta^{2}}{N}
\sum_{k=1}^{N-1}\sum_{i=1}^{k}a^{2}+
2\frac{\Delta^{2}}{N}\sum_{t=1}^{N-2}\cos\frac{2\pi nt}{N}
                            \sum_{k=1}^{N-1-t}\sum_{i=1}^{k}a^{2}\nonumber\\
&=& \frac{\Delta^{2}a^{2}}{N}\frac{N(N-1)}{2}+\frac{\Delta^{2}a^{2}}{N}\nonumber\\
&\times&\sum_{t=1}^{N-2}\big[t^{2}-(2N-1)t+N(N-1)\big]
\cos\frac{2\pi nt}{N}
\label{eq:FFTvelocity}
\end{eqnarray}
and all other terms are zero.  
The term $\sum \cos tx$ and $\sum t\cos tx$
are given as 
\begin{equation}
\sum^{N}_{t=1} \cos tx= \frac{\sin\frac{Nx}{2}}{\sin\frac{x}{2}}\cos(\frac{N+1}{2}x),
\label{eq:costx}
\end{equation}
and
\begin{equation}
\sum^{N-1}_{t=1} t\cos tx= \frac{N\sin\frac{2N-1}{2}x}{2\sin\frac{x}{2}}
-\frac{1-\cos Nx}{4\sin^{2}\frac{x}{2}}.
\label{eq:tcostx}
\end{equation}
Substituting $\frac{2\pi n}{N}$ into $x$ in Eq.\ \eqref{eq:costx}, the right hand 
side turns to zero.  The left hand side of Eq.\ \eqref{eq:costx} is 
\begin{equation}
\sum^{N}_{t=1} \cos\frac{2\pi nt}{N}= \sum^{N-2}_{t=1} \cos\frac{2\pi nt}{N}
+\cos 2\pi n+ \cos\frac{2\pi n}{N}.
\end{equation}
With the information on the both sides of Eq.\ \eqref{eq:costx}, we obtain
\begin{equation}
\sum^{N-2}_{t=1} \cos\frac{2\pi nt}{N} = -1- \cos\frac{2\pi n}{N}.
\label{eq:costx-1}
\end{equation}

Substituting $\frac{2\pi n}{N}$ into $x$ in Eq.\ \eqref{eq:tcostx}, it is then simplified as 
\begin{eqnarray}
\sum^{N-1}_{t=1} t\cos\frac{2\pi nt}{N} &=& (N-1)\cos\frac{2\pi n}{N}+
\sum^{N-2}_{t=1} t\cos\frac{2\pi nt}{N} \nonumber\\
&=& -\frac{N\sin\frac{\pi n}{N}}{2\sin\frac{\pi n}{N}} = -\frac{N}{2}.
\end{eqnarray}
Thus, we acquire 
\begin{equation}
\sum^{N-2}_{t=1} t\cos\frac{2\pi nt}{N} 
= -\frac{N}{2}-(N-1)\cos\frac{2\pi n}{N}.
\label{eq:tcostx-1}
\end{equation}

For the expression of $\sum t^{2}\cos tx$, it can be derived by differentiating 
$\sum t\sin tx$ with $x$.  The close form of $\sum t\sin tx$ is 
\begin{equation}
\sum^{N-1}_{t=1}t\sin tx = \frac{\sin Nx}{4\sin^{2}\frac{x}{2}}
-\frac{N\cos\frac{2N-1}{2}x}{2\sin\frac{x}{2}}.
\end{equation}
Differentiating it with respect to $x$ on the both sides, we get
\begin{eqnarray}
\sum^{N-1}_{t=1}t^{2}\cos tx &=& \frac{N(2N-1)\sin\frac{2N-1}{2}x}{4\sin\frac{x}{2}}\nonumber\\
&+&\frac{N[\frac{3}{2}\cos Nx+\frac{1}{2}\cos(N-1)x]}{4\sin^{2}\frac{x}{2}}\nonumber\\
&-&\frac{\sin Nx}{4\sin^{3}\frac{x}{2}}\cos\frac{x}{2}.
\end{eqnarray}
Substituting $2\pi n/N$ into $x$, it gives
\begin{eqnarray}
\sum^{N-1}_{t=1}t^{2}\cos\frac{2\pi nt}{N} &=&
(N-1)^{2}\cos\frac{2\pi n}{N}\nonumber\\
&+&\sum^{N-2}_{t=1}t^{2}\cos\frac{2\pi nt}{N}
\label{eq:t2costx-1}
\\
&=& -\frac{N(2N-1)}{4}
+\frac{N[\frac{3}{2}+\frac{1}{2}\cos\frac{2\pi n}{N}]}
{4\sin^{2}\frac{\pi n}{N}}\nonumber\\
&=& \frac{N}{2\sin^{2}\frac{\pi n}{N}}-\frac{N^{2}}{2}.
\label{eq:t2costx-2}
\end{eqnarray}
Moving the first term in Eq.\ \eqref{eq:t2costx-1} to Eq.\ \eqref{eq:t2costx-2}, we 
obtain the expression
\begin{equation}
\sum^{N-2}_{t=1}t^{2}\cos\frac{2\pi nt}{N} = -(N-1)^{2}\cos\frac{2\pi n}{N}
+\frac{N}{2\sin^{2}\frac{\pi n}{N}}-\frac{N^{2}}{2}.
\label{eq:t2costx}
\end{equation}

Substituting Eq.\ \eqref{eq:costx-1}, Eq.\ \eqref{eq:tcostx-1}, and 
Eq.\ \eqref{eq:t2costx} into Eq.\ \eqref{eq:FFTvelocity}, 
we have a compact form for the power of velocity noise
\begin{equation}
\langle|V_{n}|^{2}\rangle = \frac{\Delta^{2}a^{2}}{2\sin^{2}\frac{\pi n}{N}}.
\label{eq:powerv}
\end{equation}

Next, we consider the displacement noise of proof mass due to random acceleration 
noise.  Suppose $\{x_{i}\}$ is the time series of position noise with the initial condition of 
$x_{0} = 0$.  From Newtonian dynamics, the time series can be described by the following 
regression relationship

\begin{equation}
x_{i} = x_{i-1} + v_{i}\Delta\ \forall i: 1\sim N-1.
\label{eq:xi}
\end{equation}
From Eq.\ \eqref{eq:xi} and the initial condition we know
\begin{eqnarray}
x_{i} &=& x_{i-1} + v_{i}\Delta = x_{i-2}+(v_{i-1}+v_{i})\Delta\nonumber\\
&=& (v_{1}+\cdots+v_{i})\Delta.
\end{eqnarray}

By Eq.\ \eqref{eq:hk} the velocity can be calculated from the random acceleration noise
\begin{eqnarray}
x_{i} &=& [a_{1}+(a_{1}+a_{2})+\cdots+\sum^{i}_{k=1}a_{k}]\Delta^{2}\nonumber\\
         &=& \Delta^{2}\sum_{k=1}^{i}(i-k+1)a_{k}.
\label{eq:xi1}
\end{eqnarray}

From Eq.\ \eqref{eq:xi1} we can compute $x_{i}^{2}$ and $x_{i}x_{i+t}$
\begin{eqnarray}
x_{i}^{2} &=& \Big[\Delta^{2}\sum_{j=1}^{i}(i-j+1)a_{j}\Big]
                        \Big[\Delta^{2}\sum_{k=1}^{i}(i-k+1)a_{j}\Big]\nonumber\\
                &=& \Delta^{4}\Big[\sum_{j=1}^{i}(i-j+1)^{2}a_{j}^{2}\nonumber\\
                &+&2\sum_{j=2}^{i}\sum_{k=1}^{j-1}(i-j+1)(i-k+1)a_{j}a_{k}\Big],
\label{eq:xi2}      
\end{eqnarray}
\begin{eqnarray}
x_{i}x_{i+t} &=& \big[\Delta^{2}\sum_{j=1}^{i}(i-j+1)a_{j}\big]
                        \big[\Delta^{2}\sum_{k=1}^{i+t}(i+t-k+1)a_{j}\big]\nonumber\\
                &=& \Delta^{4}\sum_{j=1}^{i}\sum_{k=1}^{i+t}(i-j+1)(i+t-k+1)a_{j}a_{k}.
\label{eq:xixit}      
\end{eqnarray}
Substituting Eq.\ \eqref{eq:xi2} and Eq.\ \eqref{eq:xixit} into Eq.\ \eqref{eq:FFT}, 
then we can obtain the expected power of the drift

\begin{eqnarray}
\langle|X_{n}|^{2}\rangle &=& \frac{1}{N}\sum^{N-1}_{i=0}\langle x^{2}_{i}
\rangle +\frac{2}{N}\sum^{N-1}_{t=1}\sum^{N-1-t}_{i=0}\langle x_{i}x_{i+t}
\rangle\cos \frac{2\pi nt}{N}\nonumber\\
&=& \frac{\Delta^{4}a^{2}}{N}\sum_{i=1}^{N-1}
        \sum_{j=1}^{i}(i-j+1)^{2}+\frac{2\Delta^{4}a^{2}}{N}
        \sum_{t=1}^{N-2}\cos\frac{2\pi nt}{N}\nonumber\\
       &\ &\times\sum_{i=1}^{N-1-t}
        \sum_{j=1}^{i}\sum_{k=1}^{i+t}(i-j+1)(i+t-k+1)\delta_{jk}\nonumber\\
&=& \frac{\Delta^{4}a^{2}N(N^{2}-1)}{12}
      +\frac{\Delta^{4}a^{2}}{6N}\sum_{t=1}^{N-2}\cos\frac{2\pi nt}{N}\nonumber\\
      &\ &\times\left[-t^{4}+2Nt^{3}+t^{2}-2N^{3}t+N^{2}(N^{2}-1)\right].
\label{eq:powerx}           
\end{eqnarray}
\begin{figure}
\includegraphics[scale=0.31]{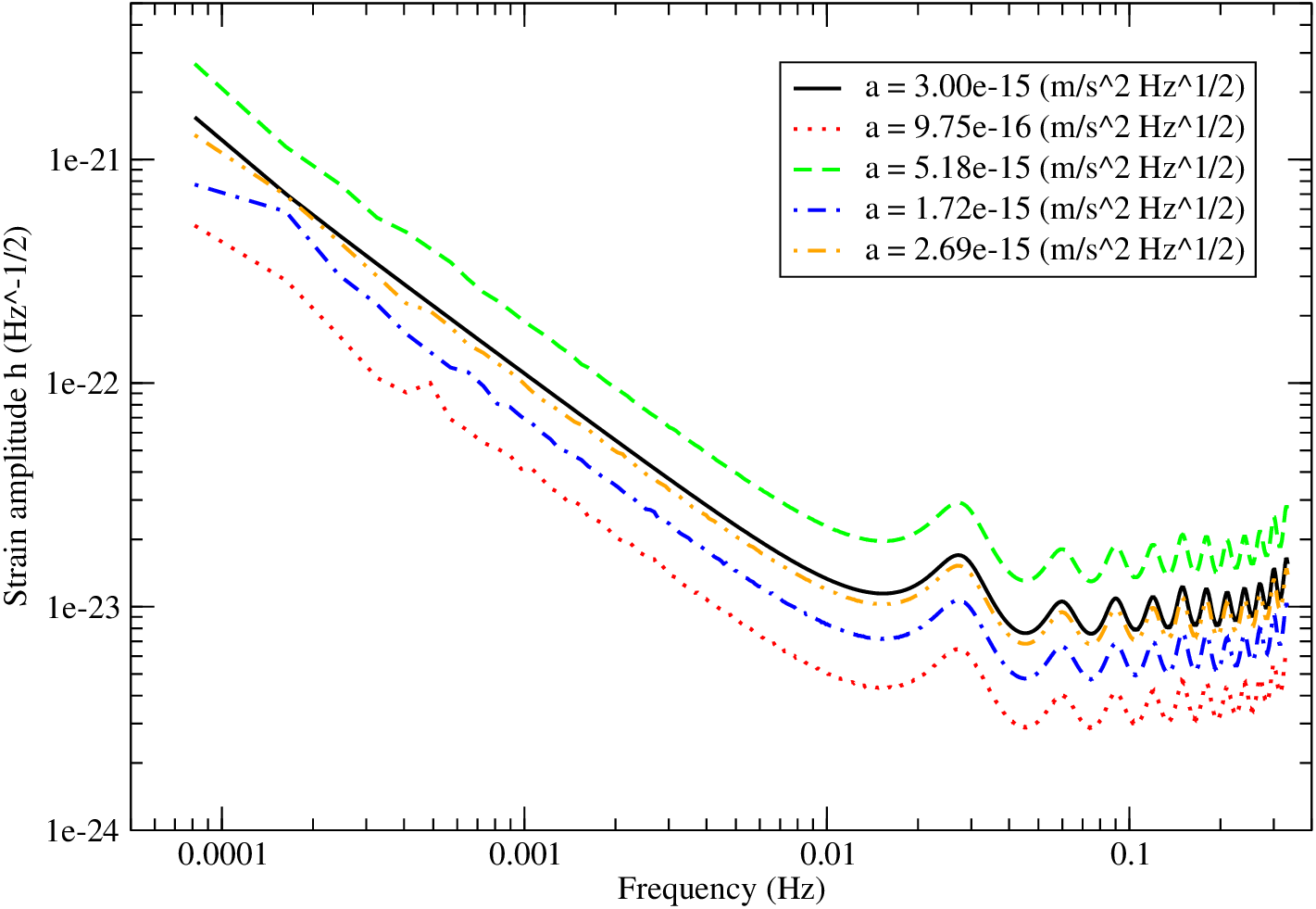}
\caption[Fourier component of acceleration-noise simulation]{
  Deconvolution of acceleration noise.  
  The black solid curve is given by Eq.\ \eqref{eq:powerx} with $a=3.00\times10^{-15}m/s^{2}$, 
  which is the value of the standard deviation used to generated time series acceleration noise.
  The other four curves denote four realisations of the deconvolution of acceleration noise, 
  respectively.  They have different estimated value of acceleration as shown in graphs.  
  }
\label{fig:Acc}
\end{figure}

We now have the expressions of $\sum_{t}t\cos\frac{2\pi nt}{N}$ and 
$\sum_{t}t^{2}\cos\frac{2\pi nt}{N}$ but we still need those of 
$\sum_{t}t^{3}\cos\frac{2\pi nt}{N}$ and $\sum_{t}t^{4}\cos\frac{2\pi nt}{N}$ 
to complete Eq.\ \eqref{eq:powerx}.  They can be obtained by differentiating 
$\sum_{t}t^{2}\sin tx$ and $\sum_{t}t^{2}\cos tx$ with respect to x once and twice, respectively, 
and then substituting $2\pi n/N$ for $x$.  Then we will get 
\begin{equation}
\sum^{N-1}_{t=1} t^{3}\cos\frac{2\pi nt}{N} = -\frac{N^{3}}{2}
+\frac{3N^{2}}{4\sin^{2}\frac{\pi n}{N}},
\label{eq:t3costx}
\end{equation}
and
\begin{equation}
\sum^{N-1}_{t=1} t^{4}\cos\frac{2\pi nt}{N} = -\frac{N^{4}}{2}
+\frac{N(N^{2}+1)}{\sin^{2}\frac{\pi n}{N}}
-\frac{3N}{2\sin^{4}\frac{\pi n}{N}}.
\label{eq:t4costx}
\end{equation}

Substituting Eq.\ \eqref{eq:tcostx}, Eq.\ \eqref{eq:t2costx}, Eq.\ \eqref{eq:t3costx}, 
and Eq.\ \eqref{eq:t4costx} into Eq.\ \eqref{eq:powerx}, we finally obtain 
\begin{equation}
\langle|X_{n}|^{2}\rangle  
=\frac{N\Delta^{4}a^{2}}{4}\left[\frac{1}{\sin^{4}\frac{\pi n}{N}}
+\frac{N^{2}-1}{3\sin^{2}\frac{\pi n}{N}}\right].
\label{eq:power-acc}
\end{equation}

There are two terms in Eq.\ \eqref{eq:power-acc} involved in the power of drifts.  One is inversely proportional to $\sin^{4}\frac{\pi n}{N}$, the other is 
$\sin^{2}\frac{\pi n}{N}$.  If n is small, we can approximate 
$\sin^{4}\frac{\pi n}{N}$ and $\sin^{2}\frac{\pi n}{N}$ by $f^{4}$ and $f^{2}$, respectively.  
The strain amplitude corresponding to these two terms can be obtained by taking square root.   
There is not only $1/f^{2}$ term, but also $1/f$ term.  Moreover, the 
strain amplitude of drifts is dominant by $1/f$ term especially in the high frequency range.  
As we have done in the analysis in the time domain, in order to reveal the true noise level, 
the power of drift must be removed in the frequency domain.  
Eq.\ \eqref{eq:power-acc} can be employed to estimate the drift power.

\section{Probability Distribution of Noise Power}
\label{sec:noise-power-prob}

The cosmic gravitational-wave background entangles with noise in time domain, forbidding the 
analysis performed.  
Hence, we intend to analyze a data set in the frequency domain through its power spectrum.  
The measurement $d(t)$ is the sum of signal $s(t)$ and Gaussian random 
noise $n(t)$ where the Gaussian noise is drawn from the distribution
\begin{equation}
P(n) = \frac{1}{\sqrt{2\pi}\sigma}\exp\Big\{-\frac{n^{2}}
{2\sigma^{2}}\Big\}.
\label{eq:prob-time-domain}
\end{equation}

\subsubsection{Fourier Amplitude of Noise}
Giving a time series of noise $\{n_{j}\}$ drawn from Eq.\ \eqref{eq:prob-time-domain},  
its Fourier amplitude is given by 
\begin{eqnarray}
\tilde{n}_{k} &=& \sum^{N-1}_{j=0} W^{kj}n_{j}\\
&=& Re\Big(n_{0}+Wn_{1}+\cdots+W^{k(N-1)}n_{N-1}\Big)\nonumber\\
&+&i\ Im\Big(n_{0}+Wn_{1}+\cdots+W^{k(N-1)}n_{N-1}\Big)\\
\label{eq:FFT-1}
&=& \sum^{N-1}_{j=0} R_{kj}n_{j} +i\sum^{N-1}_{j=0} I_{kj}n_{j}\\       
\label{eq:Xk-and-Yk}
&\equiv& \tilde{n}^{r}_{k} +i\tilde{n}^{i}_{k}
\end{eqnarray}
where $N$ is the number of noise data, and $W = \exp\{2\pi i/N\}$.  
$\tilde{n}^{r}_{k}$ and $\tilde{n}^{i}_{k}$ are the real part and imaginary part of 
$\tilde{n}_{k}$, and $R_{kj}$ and $I_{kj}$ denote the real part and imaginary part of 
$W^{kj}$, respectively.

\subsubsection{Probability of $\tilde{n}^{r}_{k}$ and $\tilde{n}^{i}_{k}$}
From the probability distribution of time series instrumental noise given by 
Eq.\ \eqref{eq:prob-time-domain}, we can derive the 
probability distribution of $\tilde{n}^{r}_{k}$ by marginalizing the probability distribution $P(\tilde{n}^{r}_{k}, n_{0}, n_{1},\cdots, n_{N-1}|I)$ over $n_{0}\cdots n_{N-1}$:
\begin{eqnarray}
P(\tilde{n}^{r}_{k}) &= & \int\cdots\int^{\infty}_{-\infty} dn_{0}dn_{1}\cdots dn_{N-1}\nonumber\\
& \times & P(\tilde{n}^{r}_{k}, n_{0}, n_{1},\cdots, n_{N-1}|\ I)
\end{eqnarray}
where $P(\tilde{n}^{r}_{k}, n_{0}, n_{1},\cdots, n_{N-1}|I)$ is the probability distribution of 
$\tilde{n}^{r}_{k}$, $n_{0},\cdots ,n_{N-1}$ being true given the background information $I$.  
From Baye's theorem, $P(\tilde{n}^{r}_{k}, n_{0}, n_{1},\cdots, n_{N-1}|I)$ can be decomposed into the product of likelihood function $P(\tilde{n}^{r}_{k}|\ n_{0},n_{1},\cdots, n_{N-1})$ which is the probability distribution of $\tilde{n}^{r}_{k}$ for a given $\{n_{0}, n_{1},\cdots, n_{N-1}\}$, and prior function $P(n_{0},n_{1},\cdots, n_{N-1})$
\begin{eqnarray}
P(\tilde{n}^{r}_{k}) &=& \int\cdots\int^{\infty}_{-\infty} dn_{0}\cdots dn_{N-1} 
P(n_{0},\cdots, n_{N-1})\nonumber\\ 
&\times& P(\tilde{n}^{r}_{k}|\ n_{0},\cdots, n_{N-1})
\label{eq:baye's}
\end{eqnarray}
Since $\tilde{n}^{r}_{k}$ is completely determined by Eq. \eqref{eq:Xk-and-Yk} 
if $n_{0}$, $n_{1}$, $\cdots$, $n_{N-1}$ are known, 
$P(\tilde{n}^{r}_{k}|\ n_{0}, n_{1},\cdots, n_{N-1})$ must satisfy
\begin{equation}
P(\tilde{n}^{r}_{k}|\ n_{0}, n_{1},\cdots, n_{N-1}) = 
\delta\Big(\tilde{n}^{r}_{k}-\sum^{N-1}_{j=0} R_{kj}n_{j}\Big)
\end{equation} 
where $\delta(x)$ is the delta function of x.  
Moreover, if $n_{0}$, $n_{1}$, $\cdots$, $n_{N-1}$ are independent random variables, 
$P(n_{0}, n_{1},\cdots, n_{N-1})$ can be written by
\begin{equation}
P(n_{0}, n_{1},\cdots, n_{N-1}) = P(n_{0})P(n_{1})\cdots P(n_{N-1}).
\end{equation}
Therefore, Eq.\ \eqref{eq:baye's} can be rewritten as
\begin{eqnarray}
P(\tilde{n}^{r}_{k}) &=& \int\cdots\int^{\infty}_{-\infty} 
\delta\Big(\tilde{n}^{r}_{k}-\sum^{N-1}_{j=1} R_{kj}n_{j}-n_{0}\Big)P(n_{0})dn_{0}\nonumber\\
&\times& P(n_{1})\cdots P(n_{N-1})\ dn_{1}\cdots dn_{N-1}.
\end{eqnarray}
Since the delta function is involved, the integration over $n_{0}$ produces $P(n_{0})$ directly where $n_{0}$ equals $\tilde{n}^{r}_{k}-\sum^{N-1}_{j=1} R_{kj}n_{j}$.  Since $P(n_{0})$ is a Gaussian distribution, we will have
\begin{widetext}
\begin{equation}
P(\tilde{n}^{r}_{k}) = \frac{1}{\sqrt{2\pi}\sigma}\int\cdots\int^{\infty}_{-\infty} 
\prod^{N-1}_{l=1}P(n_{l})\exp\bigg\{-\frac{(\tilde{n}^{r}_{k} -\sum^{N-1}_{j=1} R_{kj}n_{j})^{2}}
{2\sigma^{2}}\bigg\}dn_{1}\cdots dn_{N-1}.
\end{equation}
\end{widetext}
Combining $P(n_{1})$ with the exponential term, it becomes
\begin{widetext}
\begin{equation}
P(\tilde{n}^{r}_{k}) = \frac{1}{2\pi\sigma^{2}}\int\cdots\int^{\infty}_{-\infty}
dn_{2}\cdots dn_{N-1}\prod^{N-1}_{l=2}P(n_{l})\int^{\infty}_{-\infty} dn_{1}
\exp\bigg\{-\frac{n^{2}_{1}+\big(\tilde{n}^{r}_{k}-\sum^{N-1}_{j=2}
R_{kj}n_{j}-R_{k}n_{1}\big)^{2}}{2\sigma^{2}}\bigg\}.
\label{eq:convolution}
\end{equation}
\end{widetext}
where $P(n_{j})$ is given by Eq.\ \eqref{eq:prob-time-domain} for any integer j.  
Denoting K as the integral term over $n_{1}$ in Eq.\ \eqref{eq:convolution}, and 
$\alpha$ as $\tilde{n}^{r}_{k}-\sum^{N-1}_{j=2} R_{kj}n_{j}$, the integral can be simplified as
\begin{eqnarray}
K &=& \int^{\infty}_{-\infty} dn_{1} 
\exp\Big\{-\frac{(1+R^{2}_{k})n^{2}_{1}-2\alpha R_{k}
n_{1}+\alpha^{2}}{2\sigma^{2}}\Big\}\nonumber\\
&=& \sqrt{\frac{2\pi\sigma^{2}}{1+R^{2}_{k}}}
\exp\Big\{-\frac{\alpha^{2}}{2(1+R^{2}_{k})\sigma^{2}}\Big\}.
\label{eq:I}
\end{eqnarray}
Substituting Eq.\ \eqref{eq:I} into Eq.\ \eqref{eq:convolution} we have 
\begin{eqnarray}
&P(\tilde{n}^{r}_{k}) &=\frac{1}{\sqrt{2\pi(1+R^{2}_{k})\sigma^{2}}}
\int\cdots\int^{\infty}_{-\infty}dn_{2}\cdots dn_{N-1}\cdots\nonumber\\
& &\prod^{N-1}_{l=2}P(n_{l})\exp\Big\{-\frac
{(\tilde{n}^{r}_{k}-\sum^{N-1}_{j=2}R_{kj}n_{j})^{2}}{2(1+R^{2}_{k})\sigma^{2}}\Big\}.
\end{eqnarray}
We can integrate over from $n_{2}$ to $n_{N-1}$ with the same process used in integrating 
$n_{1}$.  Then $P(\tilde{n}^{r}_{k})$ can be written as follows
\begin{equation}
P(\tilde{n}^{r}_{k}) = \frac{1}{\sqrt{2\pi\sigma^{2}_{\tilde{n}^{r}_{k}}}}
\exp\Big\{-\frac{\tilde{n}^{r\ 2}_{k}}{2\sigma^{2}_{\tilde{n}^{r}_{k}}}\Big\}
\label{eq:prob-x-k}
\end{equation}
where $\sigma^{2}_{\tilde{n}^{r}_{k}}$ denotes $\sigma^{2}\sum^{N-1}_{j=0}R^{2}_{kj}$.
With the same steps $P(\tilde{n}^{i}_{k})$ can be found as
\begin{equation}
P(\tilde{n}^{i}_{k}) = \frac{1}{\sqrt{2\pi\sigma^{2}_{\tilde{n}^{i}_{k}}}}
\exp\Big\{-\frac{\tilde{n}^{i\ 2}_{k}}{2\sigma^{2}_{\tilde{n}^{i}_{k}}}\Big\}
\label{eq:prob-y-k}
\end{equation}
where $\sigma^{2}_{\tilde{n}^{i}_{k}}$ denotes $\sigma^{2}\sum^{N-1}_{j=0}I^{2}_{kj}$. 

\subsubsection{Probability distribution of $P_{\tilde{n}^{r}_{k}}$ and $P_{\tilde{n}^{i}_{k}}$}
The total power contained in the kth Fourier component is 
$P_{k}=P_{\tilde{n}^{r}_{k}}+P_{\tilde{n}^{i}_{k}}=\tilde{n}^{r\ 2}_{k}+\tilde{n}^{i\ 2}_{k}$ 
where $P_{\tilde{n}^{r}_{k}} \equiv \tilde{n}^{r\ 2}_{k}$ and 
$P_{\tilde{n}^{i}_{k}} \equiv \tilde{n}^{i\ 2}_{k}$.  
Now our goal is to derive the probability distribution of $P(P_{k})$ from 
Eq.\ \eqref{eq:prob-x-k} and Eq.\ \eqref{eq:prob-y-k}.  
This can be done by a series of changing variable.  At the first place the variables of 
the probability distributions were changed from $\tilde{n}^{r}_{k}$ and $\tilde{n}^{i}_{k}$ to 
$P_{\tilde{n}^{r}_{k}}$ and $P_{\tilde{n}^{i}_{k}}$, respectively, and then changed 
from $P_{\tilde{n}^{r}_{k}}$ and $P_{\tilde{n}^{i}_{k}}$ to $P_{k}$.  
Although the variables were changed, the integral of probability over entire region should be 
the same (and equal to one).  Therefore, it is known that 
\begin{eqnarray}
\int^{\infty}_{0} P(P_{\tilde{n}^{r}_{k}})\ dP_{\tilde{n}^{r}_{k}} &=& 
\int^{\infty}_{-\infty} P(\tilde{n}^{r}_{k})\ d\tilde{n}^{r}_{k}\nonumber\\
&=& 2\int^{\infty}_{0} P(\tilde{n}^{r}_{k})\ d\tilde{n}^{r}_{k}.
\label{eq:change-variable}
\end{eqnarray}
The range of $P_{\tilde{n}^{r}_{k}}$ is from zero to infinity since 
$P_{\tilde{n}^{r}_{k}}$ is positive.  
The second line is obtained from the symmetry of $P(\tilde{n}^{r}_{k})$ about zero.  
From Eq.\ \eqref{eq:change-variable} we know 
\begin{equation}
P(P_{\tilde{n}^{r}_{k}}) = 2P(\tilde{n}^{r}_{k})\ \Big |\frac{d\tilde{n}^{r}_{k}}{dP_{\tilde{n}^{r}_{k}}}\Big |
\label{eq:change-variable1}
\end{equation}
where $|\frac{d\tilde{n}^{r}_{k}}{dP_{\tilde{n}^{r}_{k}}}|$ is Jacobian.  
Since $\tilde{n}^{r}_{k} = \sqrt{P_{\tilde{n}^{r}_{k}}}$, 
the Jacobian is $\frac{1}{2}P^{-1/2}_{\tilde{n}^{r}_{k}}$.  
Substituting Eq.\ \eqref{eq:prob-x-k} and the Jacobian into Eq.\ \eqref{eq:change-variable1} 
we have
\begin{equation}
P(P_{\tilde{n}^{r}_{k}}) = \frac{1}{\sqrt{2\pi}\sigma_{\tilde{n}^{r}_{k}}}
\frac{1}{\sqrt{P_{\tilde{n}^{r}_{k}}}}\exp\Big\{-\frac{P_{\tilde{n}^{r}_{k}}}
{2\sigma^{2}_{\tilde{n}^{r}_{k}}}\Big\}.
\end{equation}
With the same steps we can derive $P(P_{\tilde{n}^{i}_{k}})$
\begin{equation}
P(P_{\tilde{n}^{i}_{k}}) = \frac{1}{\sqrt{2\pi}\sigma_{\tilde{n}^{i}_{k}}}
\frac{1}{\sqrt{P_{\tilde{n}^{i}_{k}}}}\exp\Big\{-\frac{P_{\tilde{n}^{i}_{k}}}
{2\sigma^{2}_{\tilde{n}^{i}_{k}}}\Big\}.
\end{equation}

\subsubsection{Probability Distribution of $P_{k}$}
Now we would like to know the probability distribution of $P_{k}$ for given 
$P_{\tilde{n}^{r}_{k}}$ and $P_{\tilde{n}^{i}_{k}}$.  From marginalisation we know that 
\begin{eqnarray}
P(P_{k}|\ I) &=& \iint^{\infty}_{0}dP_{\tilde{n}^{r}_{k}}dP_{\tilde{n}^{i}_{k}}\
P(P_{k}, P_{\tilde{n}^{r}_{k}},P_{\tilde{n}^{i}_{k}})\nonumber\\
&=& \iint^{\infty}_{0} dP_{\tilde{n}^{r}_{k}}dP_{\tilde{n}^{i}_{k}}\ 
P(P_{k}|\ P_{\tilde{n}^{r}_{k}},P_{\tilde{n}^{i}_{k}},I)\nonumber\\
&\times& P(P_{\tilde{n}^{r}_{k}},P_{\tilde{n}^{i}_{k}}|\ I)
\label{eq:P-Pk}
\end{eqnarray}
where the second line is given by Baye's theorem.  Since $P_{k}$ is the sum of 
$P_{\tilde{n}^{r}_{k}}$ and $P_{\tilde{n}^{i}_{k}}$, 
$P(P_{k}|\ P_{\tilde{n}^{r}_{k}}, P_{\tilde{n}^{i}_{k}}, I)$ is a delta function of 
$P_{k} - P_{\tilde{n}^{r}_{k}} - P_{\tilde{n}^{i}_{k}}$.  Because $P_{\tilde{n}^{r}_{k}}$ 
is independent of $P_{\tilde{n}^{i}_{k}}$, 
$P(P_{\tilde{n}^{r}_{k}}, P_{\tilde{n}^{i}_{k}}|\ I)$ can be decomposed as 
$P(P_{\tilde{n}^{r}_{k}})\times P(P_{\tilde{n}^{i}_{k}})$.  
Substituting $P(P_{k}|\ P_{\tilde{n}^{r}_{k}}, P_{\tilde{n}^{i}_{k}}, I)$ and 
$P(P_{\tilde{n}^{r}_{k}},P_{\tilde{n}^{i}_{k}}|\ I)$ into Eq.\ \eqref{eq:P-Pk}, and then 
integrating over $P_{\tilde{n}^{i}_{k}}$, Eq.\ \eqref{eq:P-Pk} is led to 
\begin{eqnarray}
P(P_{k}) &=& \int\int^{\infty}_{0}\delta(P_{k}-P_{\tilde{n}^{r}_{k}}-P_{\tilde{n}^{i}_{k}})\nonumber\\
& &\times P(P_{\tilde{n}^{r}_{k}})P(P_{\tilde{n}^{i}_{k}})\ dP_{\tilde{n}^{r}_{k}}dP_{\tilde{n}^{i}_{k}}.
\end{eqnarray}
Integrating the delta function over $P_{\tilde{n}^{i}_{k}}$ will give $P(P_{\tilde{n}^{r}_{k}})P(P_{k}
-P_{\tilde{n}^{i}_{k}})$, resulting in
\begin{widetext}
\begin{equation}
P(P_{k}) = \frac{1}{2\pi \sigma_{\tilde{n}^{r}_{k}}\sigma_{\tilde{n}^{i}_{k}}}
\int^{P_{k}}_{0}\ dP_{\tilde{n}^{r}_{k}}\frac{1}{\sqrt{P_{\tilde{n}^{r}_{k}}}}
\exp\bigg\{-\frac{P_{\tilde{n}^{r}_{k}}}{2\sigma^{2}_{\tilde{n}^{r}_{k}}}\bigg\}
\frac{1}{\sqrt{P_{k}-P_{\tilde{n}^{r}_{k}}}}\exp\bigg\{
-\frac{P_{k}-P_{\tilde{n}^{r}_{k}}}{2\sigma^{2}_{\tilde{n}^{i}_{k}}}\bigg\}
\end{equation}
\end{widetext}
where the upper bound of $P_{\tilde{n}^{r}_{k}}$ is subject by $P_{k}$.  Moving the exponential term involving with $P_{k}$ out of the integral, then we can obtain
\begin{widetext}
\begin{equation}
P(P_{k}) = \frac{1}{2\pi \sigma_{\tilde{n}^{r}_{k}}\sigma_{\tilde{n}^{i}_{k}}}
\exp\bigg\{-\frac{P_{k}}{2\sigma^{2}_{\tilde{n}^{i}_{k}}}\bigg\}
\int^{P_{k}}_{0}\ dP_{\tilde{n}^{r}_{k}} 
\frac{1}{\sqrt{P_{\tilde{n}^{r}_{k}}(P_{k}-P_{\tilde{n}^{r}_{k}})}}
\exp\bigg\{\frac{\sigma^{2}_{\tilde{n}^{r}_{k}}-\sigma^{2}_{\tilde{n}^{i}_{k}}}
{2\sigma^{2}_{\tilde{n}^{r}_{k}}\sigma^{2}_{\tilde{n}^{i}_{k}}}P_{\tilde{n}^{r}_{k}}
\bigg\}.
\label{eq:P-Pk-1}
\end{equation}
\end{widetext}
Expanding $\sigma^{2}_{\tilde{n}^{r}_{k}}$ and $\sigma^{2}_{\tilde{n}^{i}_{k}}$ 
Eq.\ \eqref{eq:P-Pk-1} can be simplified:
\begin{eqnarray}
\sigma^{2}_{\tilde{n}^{r}_{k}} &=& \sigma^{2} \sum^{N-1}_{j=0} R^{2}_{kj}
= \sigma^{2}\sum^{N-1}_{j=0} \cos^{2}\frac{2\pi kj}{N}\nonumber\\
&=& \frac{N}{2}\sigma^{2}+\frac{\sigma^{2}}{2}\sum^{N-1}_{j=0}\cos\frac{4\pi kj}{N},
\end{eqnarray}
and
\begin{eqnarray}
\sigma^{2}_{\tilde{n}^{i}_{k}} &=& \sigma^{2} \sum^{N-1}_{j=0} I^{2}_{kj}
= \sigma^{2}\sum^{N-1}_{j=0} \sin^{2}\frac{2\pi kj}{N}\nonumber\\
&=& \frac{N}{2}\sigma^{2}-\frac{\sigma^{2}}{2}\sum^{N-1}_{j=0}\cos\frac{4\pi kj}{N}.
\end{eqnarray}
Since the order magnitude of $\sum^{N-1}_{j=0} \cos\frac{4\pi kj}{N}$ is 1, 
$\sigma^{2}_{\tilde{n}^{r}_{k}}$ and $\sigma^{2}_{\tilde{n}^{i}_{k}}$ can be 
approximated as $N\sigma^{2}/2$ if N is large.  With this approximation we can simplify 
Eq.\ \eqref{eq:P-Pk-1} as
\begin{eqnarray}
P(P_{k}) &\approx& \frac{1}{\pi N\sigma^{2}}\exp\Big\{-\frac{P_{k}}
{N\sigma^{2}}\Big\}
\int^{P_{k}}_{0}dP_{\tilde{n}^{r}_{k}}\frac{1}{\sqrt{P_{\tilde{n}^{r}_{k}}
(P_{k}-P_{\tilde{n}^{r}_{k}})}}\nonumber\\
&=& \frac{1}{\pi N\sigma^{2}}\exp\Big\{-\frac{P_{k}}
{N\sigma^{2}}\Big\} \int^{\pi/2}_{0}\ 2d\theta\nonumber\\
&=& \frac{1}{N\sigma^{2}}\exp\Big\{-\frac{P_{k}}{N\sigma^{2}}\Big\}
\label{eq:P-Pk-21}
\end{eqnarray}
where the variable $P_{\tilde{n}^{r}_{k}}$ is changed as $P_{k}\sin^{2}\theta$.  
Since $N\sigma^{2}$ is the noise power $P_{n}$, Eq.\ \eqref{eq:P-Pk-21} can be 
rewritten in another form:
\begin{equation}
P(P_{k})=\frac{1}{P_{n}}\exp\Big\{-\frac{P_{k}}{P_{n}}\Big\}.
\label{eq:P-Pk-2}
\end{equation}
Eq.\ \eqref{eq:P-Pk-2} is the probability distribution of power in the kth Fourier 
component which we will use in data analysis.  

There are several features on $P(p_{k})$.  First, it is normalised:  
\begin{equation}
\int^{\infty}_{0} P(P_{k})\ dP_{k} = \frac{1}{P_{n}}
\int^{\infty}_{0}\ dP_{k}\ \exp\Big\{-\frac{P_{k}}{P_{n}}\Big\} = 1.
\end{equation}
Second, it is not Gaussian.  Its maximum is at $P_{k} = 0$, but its expectation value is 
the noise power $P_{n}$:
\begin{eqnarray}
\langle P_{k} \rangle &=& \frac{1}{P_{n}}\int^{\infty}_{0}
P_{k}\exp\Big\{-\frac{P_{k}}{P_{n}}\Big\}\ dP_{k}\nonumber\\
&=&\int^{\infty}_{0}\exp\Big\{-\frac{P_{k}}{P_{n}}\Big\}\ dP_{k}\nonumber\\
&=& N\sigma^{2} = P_{n}.
\label{eq:expectation-P}
\end{eqnarray}
The second line is given by integration by part.  
On the other hand, if we calculate $\langle\tilde{n}^{\ast}_{k}\tilde{n}_{k}\rangle$ 
directly from Eq.\ \eqref{eq:FFT}, it is found that
\begin{eqnarray}
\langle\tilde{n}^{\ast}_{k}\tilde{n}_{k}\rangle &=& 
\Big\langle\Big(\sum^{N-1}_{j=0}R_{kj}n_{j}\Big)^{2}+
\Big(\sum^{N-1}_{j=0}I_{kj}n_{j}\Big)^{2}\Big\rangle\nonumber\\
&=& \sum^{N-1}_{j=0} R^{2}_{kj}\langle n^{2}_{j}\rangle+\sum^{N-1}_{j=0} 
I^{2}_{kj}\langle n^{2}_{j}\rangle\nonumber\\
&=&\sum^{N-1}_{j=0}(R^{2}_{kj}+I^{2}_{kj})\sigma^{2}\nonumber\\
&=& N\sigma^{2}
\end{eqnarray}
where the second line is obtained because it is assumed that $n_{i}$ and $n_{j}$ are not 
correlated for any $i$ and $j$.  This result is consistent with Eq.\ \eqref{eq:expectation-P}.
Third, its uncertainty is $P_{n}$ as 
well:
\begin{equation}
\sigma^{2}_{P_{k}} \equiv \langle P^{2}_{k} \rangle -\langle P_{k}\rangle^{2}
\label{eq:uncertainty-def}
\end{equation}
where 
\begin{eqnarray}
\langle P^{2}_{k} \rangle &=& \frac{1}{P_{n}}\int^{\infty}_{0}
P^{2}_{k}\exp\Big\{-\frac{P_{k}}{P_{n}}\Big\}\ dP_{k}\nonumber\\
&=& 2\int^{\infty}_{0}P_{k}\exp\Big\{-\frac{P_{k}}{P_{n}}\Big\}\ dP_{k}\nonumber\\
&=& 2P_{n}\int^{\infty}_{0}\exp\Big\{-\frac{P_{k}}{P_{n}}\Big\}\ dP_{k}\nonumber\\
&=& 2P^{2}_{n}.
\label{eq:P-square}
\end{eqnarray}
Substituting Eq.\ \eqref{eq:P-square} and Eq.\ \eqref{eq:expectation-P} into 
Eq.\ \eqref{eq:uncertainty-def} we have $\sigma^{2}_{P_{k}}= P^{2}_{n}$.  
Fourth, the likelihood of $P_{k}$ located within $P_{n}\pm P_{n}$ is
\begin{equation}
\int^{2P_{n}}_{0} P(P_{k})\ dP_{k} = 1- e^{-2} = 86.47\%.
\end{equation} 
\begin{figure}
\includegraphics[scale=0.31]{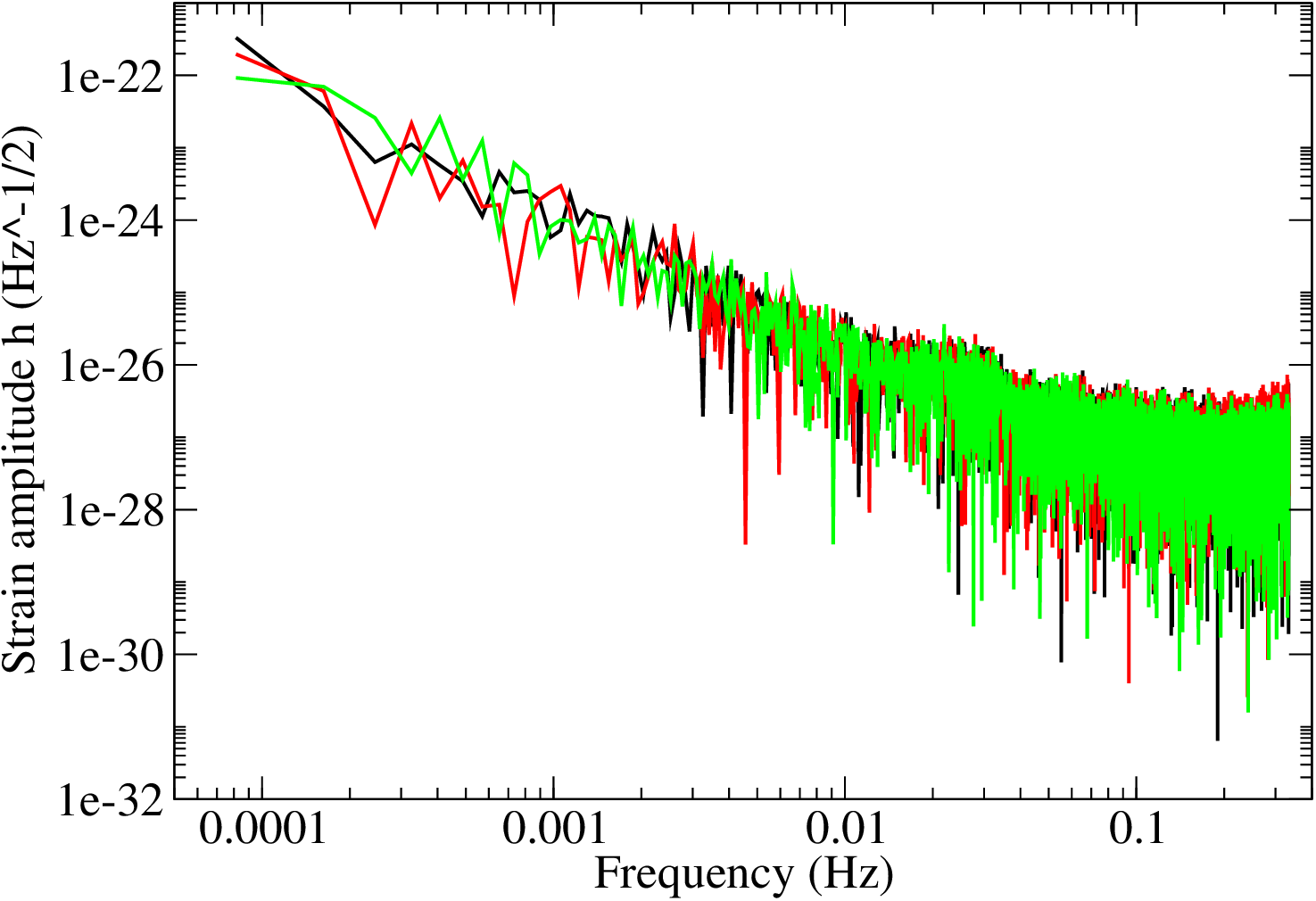}
\caption[The residuals of eight acceleration noise after de-trending]{
  The deconvolution of residuals of three realisations of acceleration noise after de-trending are displayed 
  in different curves.  The estimated acceleration of black, red, and green curves are 
  $9.75\times 10^{-16}$, $5.18\times 10^{-15}$, 
  and $1.72\times 10^{-15}\ m/s^{2}Hz^{1/2}$, respectively.  
  The residuals were converted from the observation time of 12288 sec 
  to 1-yr observation.  
  }
\label{fig:Acc-residual}
\end{figure}

\section{Data Analysis}
\label{sec:analysis}

The parameter estimation algorithm is built upon the Bayesian statistics.  \cite{bayesian}
Baye's theorem allows us to decompose the probability density function of hypothesis into 
likelihood function and prior, which are easier to assign.  
The main challenge of the algorithm is to complete estimation within a reasonable time.  
For instance, even though just ten values were tried for each parameter, $10^{10}$ trial parameter sets are necessary to find the best estimate if the model contains 10 parameters.  
To solve this, Markov Chain Monte Carlo (MCMC) method \cite{MCMC} 
was adapted.  Furthermore, the simulated annealing \cite{KGV1983, Cerny1985} 
is applied to speed up the search of the Markov Chain.  
With the Markov Chain 
Monte Carlo method, we can sample parameters from the likelihood function.  If the number 
of samples is plenty, the distribution of the parameter shall be close to the posterior.  Then 
the value of parameters, the uncertainties, and the correlations can be estimated directly 
from the samples.  

The acceptance of the candidate state is conditional on its relative probability 
to the current state $P(\mathbf{x}^{(n+1)}|data, I)/P(\mathbf{x}^{(n)}|data, I)$.  
With Baye's theorem, 
we can expand $P(\mathbf{x}^{(n+1)}|data, I)$ and $P(\mathbf{x}^{(n)}|data, I)$.  
Since the same model and the same data set are used, the prior and the evidence will be 
cancelled out.  
The relative probability is then reduced to their likelihood ratio
\begin{equation}
\frac{P(\mathbf{x}^{(n+1)}|data, I)}{P(\mathbf{x}^{(n)}|data, I)}
=\frac{P(data|\mathbf{x}^{(n+1)}, I)}{P(data|\mathbf{x}^{(n)}, I)}.
\label{eq:likelihood-ratio}
\end{equation}
Substituting Eq.\ \eqref{eq:P-Pk-2} into Eq.\ \eqref{eq:likelihood-ratio} 
the likelihood ratio can be expressed as 
\begin{equation}
\frac{P(\mathbf{x}^{(n+1)}|data, I)}{P(\mathbf{x}^{(n)}|data, I)}=\Pi_{i}
\frac{\exp\Big\{-\frac{1}{P_{n}}\big(p^{d}_{i}-p^{s}_{i}(\mathbf{x}^{(n+1)})
\big)\Big\}}{\exp\Big\{-\frac{1}{P_{n}}\big(p^{d}_{i}-p^{s}_{i}(\mathbf{x}^{(n)})
\big)\Big\}}
\label{eq:probability-ratio}
\end{equation}
where $p^{d}_{i}$ is the power in the $i$th data, 
and $p^{s}_{i}(\mathbf{x}^{(n+1)})$ and $p^{s}_{i}(\mathbf{x}^{(n)})$ are the 
signal power given by the state $\mathbf{x}^{(n+1)}$ and $\mathbf{x}^{(n)}$, 
respectively.  With this relative probability we can compute the transition probability 
\begin{equation}
A((\mathbf{x}^{(n+1)}),(\mathbf{x}^{(n)}))
=min\Bigg(1,\frac{P(\mathbf{x}^{(n+1)}|data, I)}{P(\mathbf{x}^{(n)}|data, I)}\Bigg)
\label{eq:transition-prob}
\end{equation}
to determine whether the candidate state should be accepted.

There are two concerns about the convergence of chain.  
One is that the chain may converge very slow.  The other is that the chain may not converge 
as the sampling is terminated.  To deal with the first one, the simulated annealing is applied.  
This is a dynamic way to change the moves between the samples.  It encourages bold moves 
in the beginning of the sampling to prevent the chain from getting stuck in local minimums.  
After this burn-in phase, the moves will be adapted to be conservative to speed up the 
sampling.  For the second one, a diagnostic to monitor the convergence of chain is necessary.  
According to the way of monitoring, the diagnostics were classified as qualitative (graphical) or quantitative.  Some of them generate a very long chain to do monitoring, and the others generate multiple shorter chains to proceed. 
The Gelman \& Rubin's method \cite{Gelman-Rubin} is used as a diagnostic of the convergence of chain.  It is used in various fields, such as Cosmic Microwave Background data analysis \cite{Gelman-Rubin-CMB}.  
Gelman \& Rubin's method observes the convergence with multiple-chain approach.  
The idea is that with-in chain variance and between-chain variance shall be very close if the chains are converged; otherwise, the between-chain variance shall be larger than with-in chain variance.  
Cowles and Carlin have reviewed various diagnostics of convergence \cite{Convergence-Review}.

Combining the Bayes's theorem, the Markov Chain Monte Carlo method, the simulated 
annealing method, and the Gelman \& Rubin's method, the detail process is as follows.  
First, draw number of starting points of chains from a uniform distribution.  
The required number of chains is 10 times the number of parameters in the model.  
Second, apply simulated annealing to generate candidate states.  Third, use 
Eq.\ \eqref{eq:P-Pk-2} to calculate the likelihood ratio between two states as 
transition probability $P$.  If $P$ is larger 
than 1, then the candidate state is accepted.  If $P$ is smaller than 1, we generate a 
random number $r$ from a uniform distribution for $0\le r < 1$.  If $r \le P$, then the 
candidate state is still accepted; otherwise, the candidate state is rejected, and we generate 
a new one to repeat the process.  Fourth, use the Gelman \& Rubin's method to calculate the 
factor $\hat{R}$ to monitor the convergence.  If $\hat{R} > 1.01$, this indicates that the 
chains may not be converged, and more samples are required.  If $\hat{R} < 1.01$, it is 
suggested that the chains are converged, and we use the second half of the chains to estimate 
the expectation value 
%expectation value
\begin{equation}
\langle x^{i}\rangle = \frac{1}{N}\sum^{N}_{n=1} \tilde{x}^{i}_{n}, 
\end{equation}
uncertainties
%uncertianty
\begin{equation}
\sigma_{x^{i}}= \sqrt{\frac{1}{N-1}\sum^{N}_{n=1}(\tilde{x}^{i}_{n})^{2}
-\frac{N}{N-1}\langle x^{i}\rangle^{2}},
\end{equation}
and correlations
%correlation
\begin{equation}
cor(x^{i},x^{j})=\frac{\sum^{N}_{n=1}\tilde{x}^{i}_{n}\tilde{x}^{j}_{n}
-N\langle x^{i}\rangle\langle x^{j}\rangle}
{(N-1)\sigma_{x^{i}}\sigma_{x^{j}}}
\end{equation}
of parameter set $\{x^{i}\}$ where $i=1,2,\cdots,m$ and $N$ is the length of the second 
half of the chains $\tilde{x}^{i}$.  

We simulated 24 realisations of deconvolution of the random acceleration.  
The deconvolution is given by
\begin{equation}
h_{a} = \sqrt{\frac{P_{a}(f)}{R(f)}}
\end{equation}
where $P_{a}(f)$ is the power of random acceleration given by Eq.\ \eqref{eq:power-acc}, 
and $R(f)$ is the LISA transfer function.  
The value of the standard deviation used in the Gaussian distribution for generating 
acceleration noise is $3.00\times10^{-15}\ m/s^{2}\sqrt{Hz}$.  
Fig. \ref{fig:Acc} shows four of realisations.  
The black solid curve indicates the deconvolution $h_{a}$ with this magnitude.  
As revealed in the figure, all realisations have different amplitudes but share the same 
pattern.  
The amplitude of acceleration noise spectrum 
is proportional to the final trajectory discrepancy of proof-mass from its free-fall track.  
This suggests that the coefficient $a$ in Eq.\ \eqref{eq:power-acc} is the `average' 
acceleration corresponding to each realisation, and it 
inherits the random nature of time series acceleration noise.   
The removal of acceleration noise aims to find a value for $a$ to provide the best description of 
data with Eq.\ \eqref{eq:power-acc}.
 
\begin{figure}
\includegraphics[scale=0.31]{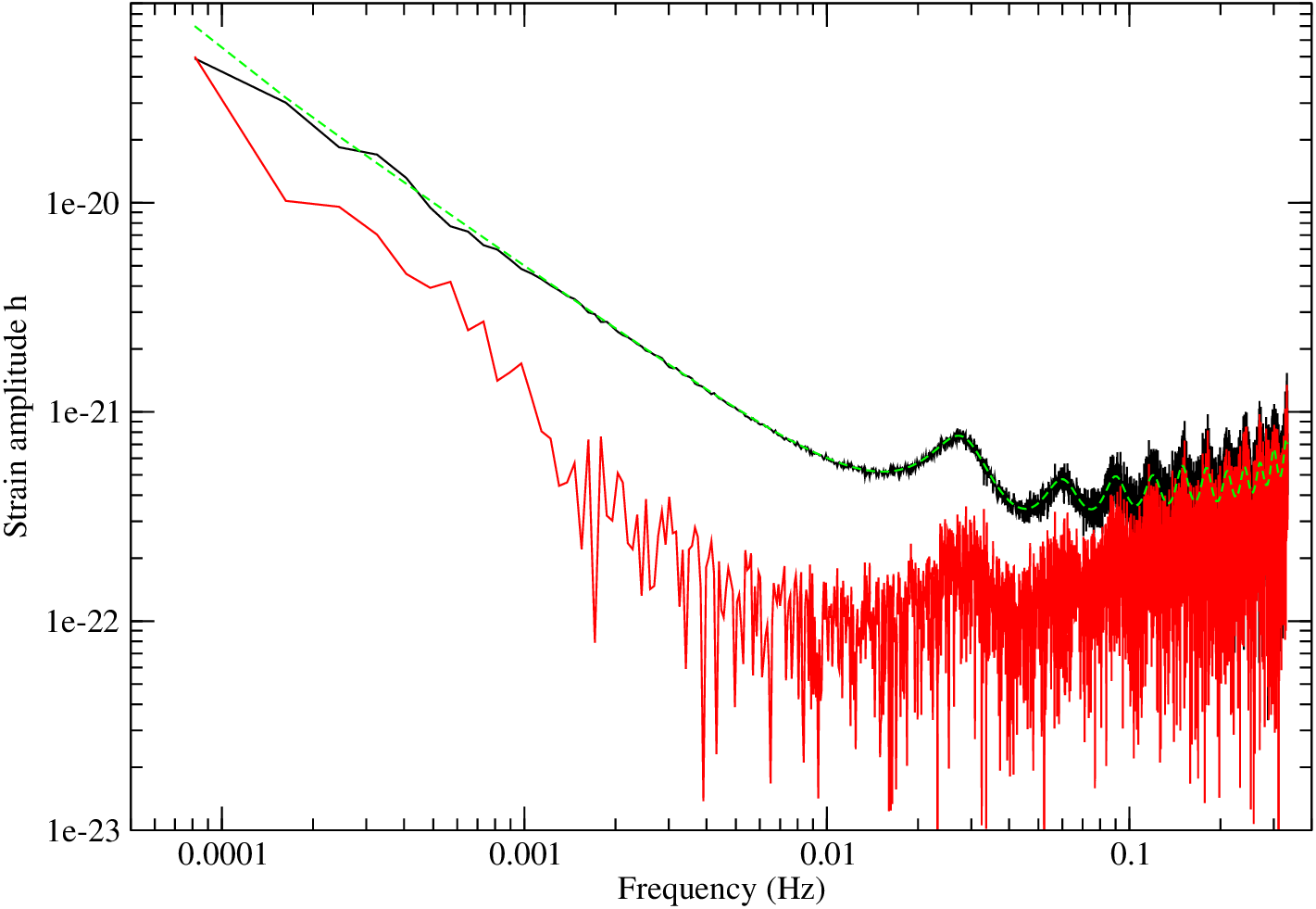}
\caption[Data reduction of instrumental noise]{
  The black curve is the strain of the sum of shot noise and drift trend induced from the 
  random acceleration.  
  The green dash curve represents the estimation of the drift trend.  
  The bottom red curve is the residuals after removal, regarding as the instrumental noise.  
  The observation time is 12288 sec.
  }
\label{fig:Ins-residual}
\end{figure}

The magnitude of a is estimated through the parameter estimation algorithm.  
The samples of a are drawn from Eq.\ \eqref{eq:transition-prob} where $p^{s}_{i}$ is 
given by 
\begin{equation}
P^{s}(f_{i}) = \frac{N\Delta^{4}a^{2}}{4}\left[\frac{1}{\sin^{4}\frac{\pi i}{N}}
+\frac{N^{2}-1}{3\sin^{2}\frac{\pi i}{N}}\right], 
\label{eq:ins}
\end{equation}
and noise power $P_{n}(f)$ is given by 
\begin{equation}
P_{n}(f)=\frac{1}{12288L^{2}R(f)}\Big\{S_{s}+\frac{S_{a}}{f^{4}}\Big\}
\end{equation}
where $L$ is the arm-length $5\times 10^{9}\ m$, 12288 is our observation time in second, 
$R(f)$ is the LISA transfer function, 
$S_{s}$ is the shot noise spectral density $1.04\times 10^{-22}\ m^{2}/Hz$, and 
$S_{a}$ is the acceleration noise spectral density $9\times10^{-30}\ m^{2}/s^{4}\ Hz$.  
Although the best estimates of $a$ are all different for each realisation, 
the noise levels are the same as Fig. \ref{fig:Acc-residual} shows.  

The raw data of instrumental noise, as presented by black curve in Fig. \ref{fig:Ins-residual}, 
is sum of drift trend caused by random acceleration and shot noise.   
The estimation of the drift trend is displayed by the green dash curve in Fig. \ref{fig:Ins-residual}.
The residual after the removal is 
shown by the red curve in Fig. \ref{fig:Ins-residual}.  In the frequency range below 
$1\ mHz$ where acceleration noise is dominant, the residuals decrease as $1/f^{2}$ as 
expected from acceleration noise.  
In the frequency range above $10\ mHz$ where shot noise is stronger, 
the residuals roughly increases as $f$.  
Between $1\sim 10\ mHz$ where shot noise and acceleration noise are comparable, 
the lowest region of the noise is around $5\sim 10\ mHz$.

\section{Conclusion}
\label{sec:conclusion}

In Sec.\ \ref{sec:time-domain} we have illustrated the removal of the drift trend of LISA proof mass in the time domain.  By using a quadratic function to fit the data, the trend can be estimated and be removed from the data.  Converting the time span of the data to one year observation, the frequency dependance and magnitude of the cleaned data match the low frequency part of LISA sensitivity curve.  
In addition, a cubit function was used to fit the trend as well, but it did not give a better fitting, 
indicating that the the instrumental noise in the low frequency band as shown in the LISA sensitivity curve is intrinsic.

Since the cosmological sources are randomly distributed across the sky, the emitted 
gravitational waves will form a continuum and be mixed with the LISA instrumental noise in the time domain.  
In order to separate the background from data, the power spectrum of the drift trend of the proof mass shall be found in the frequency domain.  
To achieve this, firstly we have re-formulated the Fourier power spectrum in Sec.\ \ref{sec:ins-noise-power}.  
With this new representation, the product of time series data $n_{i}n_{j}$ in the Fourier transform is separated into autocorrelation part $n^{2}_{i}$ and cross-correlation part $n_{i}n_{j}$ where $i\not=j$.  
The advantage of this representation is that the ensemble average of the cross-correlation part will 
vanish when we deal with purely noisy data.  

Next, we have applied the new representation to derive the expected power spectrum of the drift.  
It is found that the strain amplitude of shot noise is white, 
and that of velocity noise induced from random acceleration follows $\sim 1/f$ as expected.  
As for the amplitude of displacement noise caused from the acceleration, it is thought to be 
proportional to inverse square of frequency, resulting from integrating time by part twice.  
However, we realized that it depends not only on a term associated with $\sim 1/f^{2}$, 
but also on a term proportional to $1/f$.  
Moreover, the $1/f$ term is the dominant component.  

It is known that the time series data of the drift is subject to Gaussian noise, but we cannot sure that its counterpart in the frequency domain is subject to Gaussian noise as well.  In Sec.\ \ref{sec:noise-power-prob} we have derived the probability distribution which the power of the drift in the frequency domain is subject.  The probability distribution is exponential.  Some characters of the distribution, such as the mean and uncertainty, have been given as well.

In Sec.\ \ref{sec:analysis} the algorithm for the data analysis has been described.  
The algorithm was built upon the Bayesian statistics, which was collaborated with a Markov 
Chain Monte Carlo method to enhance the efficiency of the analysis.  
Simulated annealing was employed to encourage Markov Chains to explore entire parameter space.  
The Gelman \& Rubin method (1992) was chosen as a diagnostic for the convergence of the chains to confirm all statistical results being reliable.  

We have employed the algorithm to analyze 24 realizations of drift trend lasting 12288 sec.  
It is found that the frequency dependence of the strain given by the acceleration noise is $\sim 1/f^{2}$.  
We convert our results to a data set of 1-year observation time by multiplying a factor of $\sqrt{12288\sec/1-yr \ in\ sec}$, 
giving that the strain amplitude induced from acceleration noise is around $5\times 10^{-24}$ at $1\ mHz$,
which agrees with the acceleration noise in the figure 4.3 in \cite{prephaseA}.  

In this paper we have demonstrated an approach to clean the power of the drift induced from the random accelerations on LISA proof mass in the frequency domain.  
The approach can be applied to other space-borne interferometers as well if charges on their proof masses cannot be perfectly cancelled.  
We have shown that the LISA sensitivity can be recovered with this approach.  
This approach allows us to construct a more complicated algorithm to detect stochastic gravitational-wave background in the LISA data stream.  

\bibliography{acceleration_noise}% Produces the bibliography via BibTeX.

\end{document}